%% file: main.tex
\documentclass[twocolumn]{article}

\usepackage{subcaption}
\usepackage{graphicx}
\usepackage{color}
\usepackage{soul} 
\usepackage{multirow}
\usepackage{appendix}
\usepackage{subcaption} 
\usepackage{ragged2e}
\usepackage[hyphens]{url}
\usepackage{booktabs}
\usepackage{amsmath}
\usepackage{amssymb}
\usepackage{inputenc}
\usepackage{array}
\usepackage{epsfig}
\usepackage{amsfonts}
\usepackage{geometry}

\DeclareUnicodeCharacter{2212}{-}

\newcolumntype{M}[1]{>{\centering\arraybackslash}m{#1}}

\providecommand{\keywords}[1]
{
  \small	
  \textbf{\textit{Keywords---}} #1
}

\title{\textbf{Modelling Urban Dynamics with Multi-Modal Graph Convolutional Networks}}

\author{
    \textbf{Krittika D'Silva}\\
    \texttt{Computer Laboratory,}\\
    \texttt{University of Cambridge}\\
    \texttt{krittika.dsilva@cl.cam.ac.uk}

\and
    \textbf{Jordan Cambe}\footnote{Work was performed while a visitor at the University of Cambridge. Complete affiliation:
    Univ Lyon, ENS de Lyon, UCB Lyon 1, CNRS, Laboratoire de Physique, F-69342 Lyon, France;
    Institut Rh\^{o}nalpin des Systemes Complexes, IXXI, F-69342 Lyon, France}\\
    \texttt{Laboratoire de Physique,}\\
    \texttt{ENS de Lyon}\\
    \texttt{jordan.cambe@gmail.com}

\and
    \textbf{Anastasios Noulas} \\ 
    \texttt{New York University}\\
    \texttt{tnoulas@gmail.com}

\and   
    \textbf{Cecilia Mascolo} \\
    \texttt{Computer Laboratory,}\\ 
    \texttt{University of Cambridge}\\
    \texttt{cm542@cam.ac.uk}

\and
    \textbf{Adam Waksman} \\
    \texttt{Foursquare Labs}\\
    \texttt{awaksman@foursquare.com}
}

\date{}

\begin{document}

\maketitle

\input{1_abstract.tex} 


\section{Introduction}
\input{2_intro.tex} 

\section{Related Work}
\input{3_related.tex}

\section{Dataset Description}
\input{4_dataset.tex}
\label{sec:data}

\section{Urban Activity Networks}
\label{sec:network}
\input{5_urbanactivitynets.tex}

\section{Methodology}
\label{sec:methodology}
\input{6_model_analysis.tex}

\section{Results \& Discussion} 
\input{7_results.tex}

\section{Conclusion And Future Work}
\label{sec:conclusion}
\input{8_conclusion.tex}

\section{Acknowledgements}
We thank Foursquare for supporting this research by providing the dataset used in the analysis. \\
Jordan Cambe would like to thank the GdR-ISIS and the ED PHAST (52) for supporting his work in the University of Cambridge. \\
Krittika D'Silva's work was supported through the Gates Cambridge Trust.

\bibliographystyle{abbrv}
\bibliography{main} 

\end{document}

%% file: 1_abstract.tex
\begin{abstract}
Modelling the dynamics of urban venues is a challenging task as it is multifaceted in nature. Demand is a function of many complex and nonlinear features such as neighborhood composition, real-time events, and seasonality. Recent advances in Graph Convolutional Networks (GCNs) have had promising results as they build a graphical representation of a system and harness the potential of deep learning architectures. However, there has been limited work using GCNs in a temporal setting to model dynamic dependencies of the network. Further, within the context of urban environments, there has been no prior work using dynamic GCNs to support venue demand analysis and prediction. In this paper, we propose a novel deep learning framework which aims to better model the popularity and growth of urban venues. Using a longitudinal dataset from location technology platform Foursquare, we model individual venues and venue types across London and Paris. First, representing cities as connected networks of venues, we quantify their structure and characterise their dynamics over time. We note a strong community structure emerging in these retail networks, an observation that highlights the interplay of cooperative and competitive forces that emerge in local ecosystems of retail establishments. 
Next, we present our deep learning architecture which integrates both spatial and topological features into a temporal model which predicts the demand of a venue at the subsequent time-step. Our experiments demonstrate that our model can learn spatio-temporal trends of venue demand and consistently outperform baseline models. Relative to state-of-the-art deep learning models, our model reduces the RSME by $\approx$ 28\% in London and $\approx$ 13\% in Paris. Our approach highlights the power of complex network measures and GCNs in building prediction models for urban environments. The model has numerous applications within the retail sector to better model venue demand and growth. More broadly, the methodology and results can support policymakers, business owners, and urban planners in the development of models to characterize and predict changes in urban settings.
\end{abstract}

%% file: 2_intro.tex
Predicting venue demand has long been an active area of research because of its inherent value for numerous stakeholders. Location is known to be highly influential in the success of a new business opening in a city. Where a business is positioned across the urban plane not only determines its reach by clienteles of relevant demographics, but more critically, it determines its exposure to a local ecosystem of businesses who strive to increase their own share in a local market. The types of businesses and brands that are present in an urban neighborhood in particular has been shown to play a vital role in determining whether a new retail facility will grow and blossom, or instead whether it will become a sterile investment and eventually close~\cite{jensen2}. Competition is nonetheless only one determinant in retail success. How a local business establishes a \textit{cooperative network} with other places in its vicinity has been shown to also play a decisive role in its sales growth~\cite{daggitt}. Local businesses can complement each other by exchanging customer flows with regards to activities that succeed each other (e.g. going to a bar after dining at a restaurant), or through the formation of urban enclaves of similar local businesses that give rise to characteristic identities that then become recognisable by urban dwellers. A classic example of the latter is the presence of many Chinese restaurants in a Chinatown~\cite{zhou2010chinatown}.

It is therefore natural to hypothesize that the rise of a business lies on the complex interplay between cooperation and competition that manifests in a local area. Modelling these cooperative and competitive forces in a city remains, however, a major challenge. Today's cities change rapidly driven by urban migration and phenomena such as gentrification as well as large urban development projects, which can lead to shops opening and closing at increasing rates. In 2011, the \textit{fail rate} of restaurants in certain cities, such as New York, was as high as 80\% \footnote{\raggedright\url{https://www.businessinsider.com/new-york-restaurants-fail-rate-2011-8}} with some businesses closing in only a matter of months. 
A similar picture has been reported for high street retailers in the United Kingdom with part of the crisis being attributed to the increasing dominance of online retailers
\footnote{\raggedright\url{https://www.theguardian.com/cities/ng-interactive/2019/jan/30/high-street-crisis-town-centres-lose-8-of-shops-in-five-years}}.  
Data generated in location technology platforms by mobile users who navigate the city provides a unique opportunity to respond to the aforementioned challenges. In addition to providing quick updates, in almost real time, on venue demand in cities - thus accurately reflecting the visitation patterns of local businesses in a given area - they offer a view on urban mobility flows between areas and places at fine spatial and temporal scales. The ability to describe these two dimensions of urban activity - places and mobility - paves the way for predicting venue demand at subsequent time-steps. In this work, we harness this opportunity, building on a longitudinal dataset by Foursquare that describes mobility interactions between places in London and Paris. There has been limited prior work using temporal Graph Convolutional Networks (GCNs) to model trends in urban areas. Our work is a first step in harnessing the power of deep learning in conjunction with graph attributes to predict future growth trends for venues in a given area. Our contributions are summarized in more detail in the following: 
\begin{itemize}
\item{\textbf{Detecting patterns of cooperation in urban activity networks:} We model businesses in a city as a connected network of nodes belonging to different activity types. We examine the properties of these networks spatially and temporally. With respect to null network models, we observe higher clustering coefficient, higher modularity and lower closeness centrality scores which are indicators of strong tendencies for local businesses to cluster and form collaborative communities that exchange customer flows. Strong community structure emerges in these local retail networks and cooperative effects appear naturally, reflecting interactions of nodes within and across communities. 
In numerical terms, the modularity of urban activity networks is $\approx 0.6$  relative to the corresponding null models with $\approx 0.15$.}

\item{\textbf{Novel temporal GCN architecture which combines topological and spatial structure:} We describe our architecture which incorporates both spatial and topological features into a Graph Convolutional Network. We begin by visualizing the temporality of our data and how it can be used to build a network representation. We then detail our architecture which has the potential to be broadly applicable across numerous domains.}

\item{\textbf{First temporal GCN applied to urban dynamic prediction:} We employ our spatio-topological temporal model to predict the demand of a given venue at the subsequent month. Our model has a high predictive power over baselines, reducing the RSME of demand prediction by $\approx$ 28\% in London and  $\approx$ 13\% in Paris relative to an LSTM. We consistently outperform baseline models across both metrics and both cities of interest. We further show that our model converges more quickly suggesting it is able to quickly learn trends in venue popularity using both topological and spatial features. Our approach shows how complex networks can be used in dynamic deep learning models. Further, it has the potential for broad applications in urban planning and retail through predictions of future growth trends.
}
\end{itemize}

Our results are especially important in a digital age with shifting customer preferences as physical business are forced to adapt to remain competitive. Our methodology can enable a better understanding of interactions within local retail ecosystems.
Modern data and methods, such as those employed in the present work, not only can allow for monitoring these phenomena at scale, but also offer novel opportunities for retail facility owners to assess future demand trends though  location-based analytics. Similar methods can be applied beyond the scope of the retail sector we study here, namely for urban planning and innovation e.g. through assessing future demand trends of transport hubs, leisure and social centers or health and sanitation facilities in city neighborhoods. 
\begin{figure*}
\centering
\begin{minipage}{.5\textwidth}
  \centering
  \includegraphics[width=8.9cm]{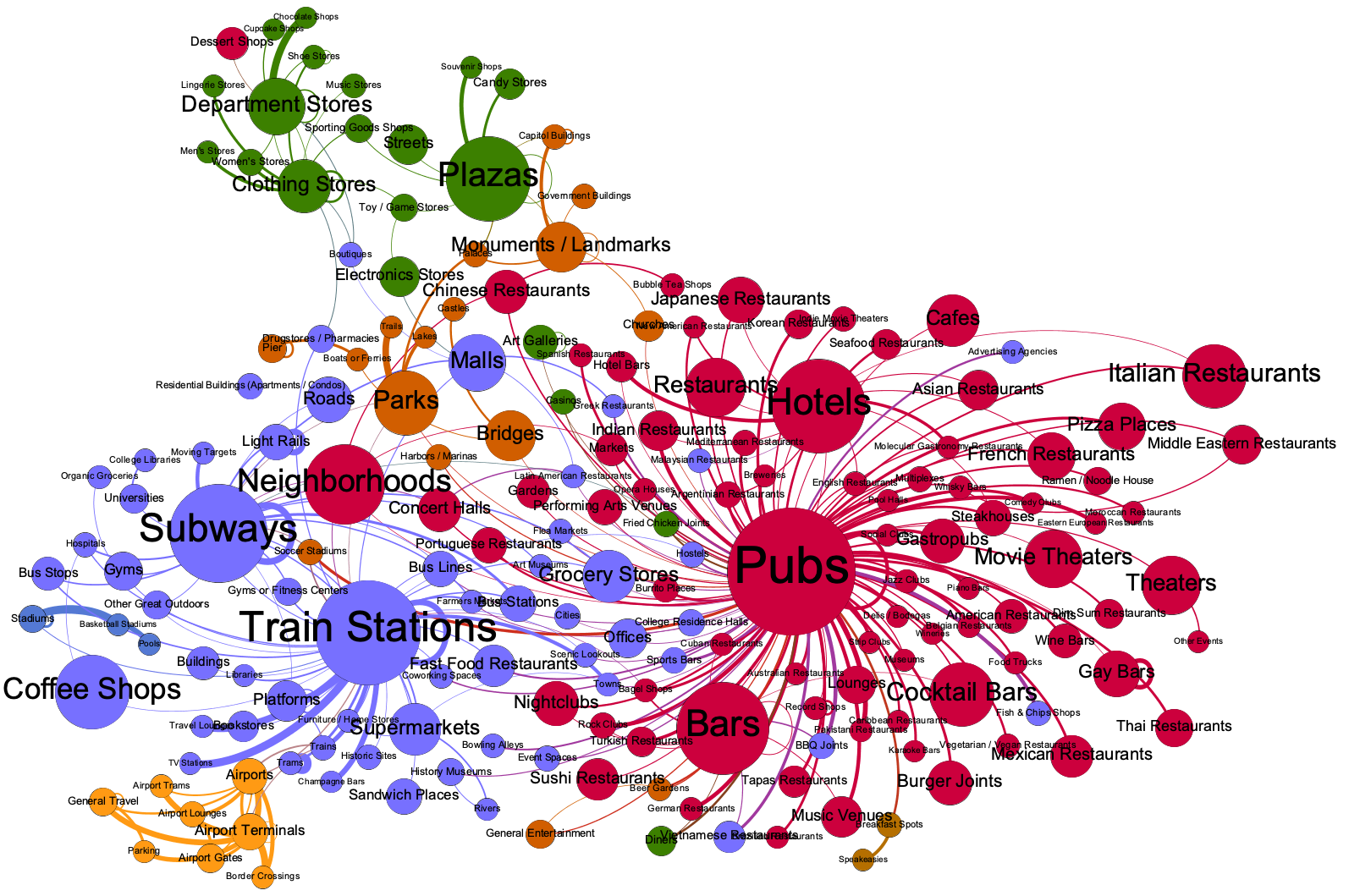}  
\end{minipage}%
\begin{minipage}{.5\textwidth}
  \centering
  \includegraphics[width=8.9cm]{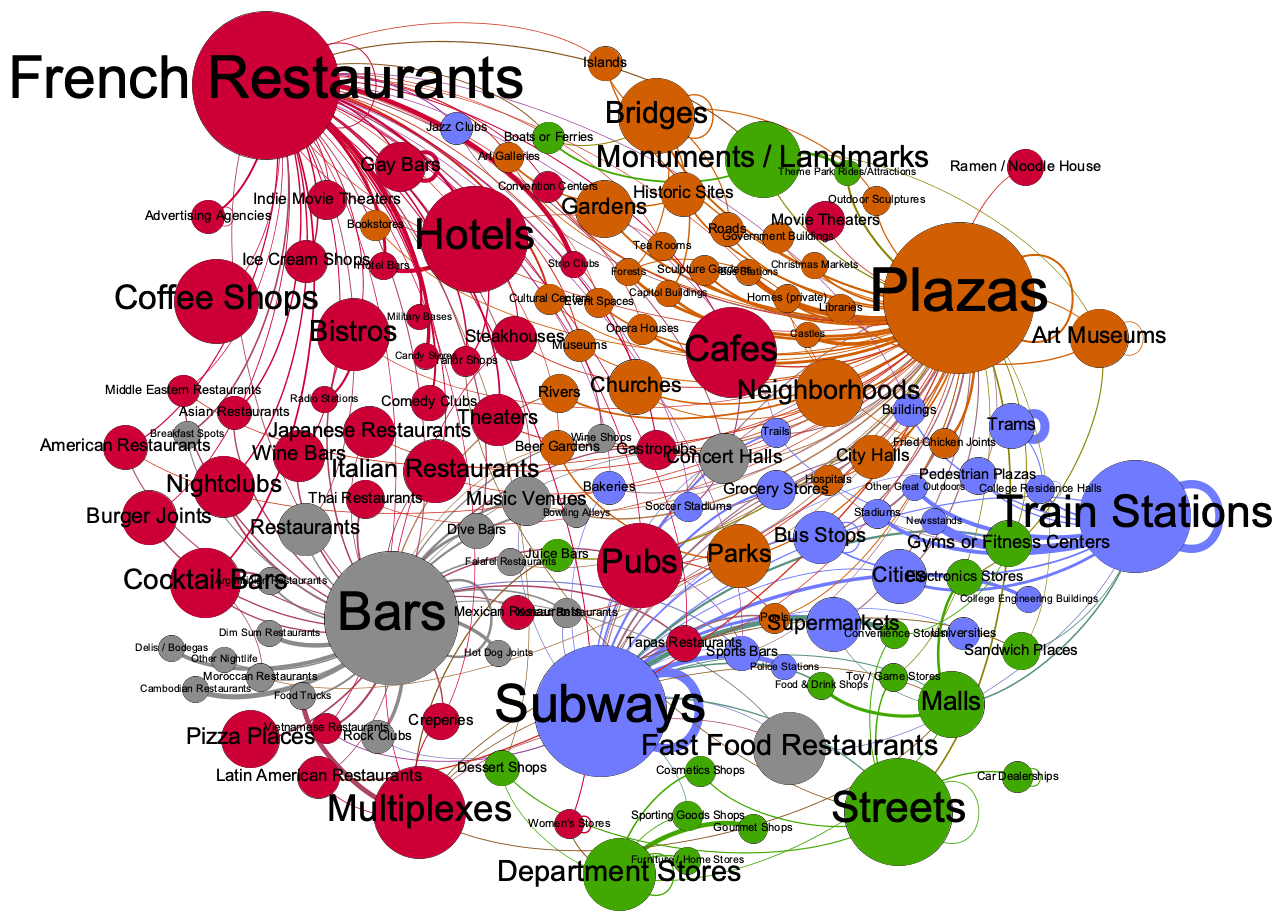}
\end{minipage}
    \caption{Network visualization of categories during the evening in London (left) and Paris (right). Different colors represent different Louvain communities ~\cite{blondel_2008}. We see clusters of travel and transport (blue), nightlife (red), and shopping activities (green).}
    \label{fig:networkViz}
\end{figure*}

%% file: 3_related.tex
Understanding retail ecosystems and determining the optimal location for a business to open have been long been questions in operations research and spatial economics~\cite{ghosh1983formulating,eiselt1989competitive}. Compared to modern approaches, these methods were characterized by static datasets informing on population distribution across geographies, tracked through census surveys and the extraction of retail catchment areas through spatial optimization methods~\cite{applebaum1966methods}.
Gravity models on population location and mobility later became a common approach for site placement of new brands~\cite{gibson1972retail}.

By the early 90s many major brands were already developing their expansion and growth strategy using quantitative methods in a manner that reflected the demographics and preferences of local populations~\cite{cheng2004exploring}. Multi-national corporations like Starbucks, Mc Donald's and 7-eleven fell into this category in their search for international expansion~\cite{sternquist1997international}.

The availability of spatio-temporally granular urban datasets and the popularization of spatial analysis methods in the past decade led to a new generation of approaches to quantify retail success in cities. In this line, network-based approaches have been proposed to understand the retail survival of local businesses through quality assessment on the interactions of urban activities locally~\cite{jensen2}. In addition to networks of places, street network analysis emerged as an alternative medium to understand customer flows in cities, with various network centrality being proposed as a proxy to understand urban economic activities~\cite{crucitti2006centrality,porta2012street}.
More recently, machine learning and optimization methods have been introduced to solve location optimization problems in the urban domain, focusing not only on retail store optimization~\cite{karamshuk2013geo} but also real estate ranking~\cite{fu2014sparse} amongst other applications. Location technology platforms such as Foursquare opened the window of opportunity for customer mobility patterns to be studied at fine spatio-temporal scales~\cite{d'silva1, d'silva2} and moreover, semantic annotations on places presented direct knowledge on the types of urban activities that emerge geographically and led to works that allowed for the tracking and comparing urban growth patterns at global scale~\cite{daggitt}. Additionally, the authors in~\cite{hidalgo} study co-location patterns of urban activities in Boston and subsequently recommend areas where certain types of activities may be missing. 

Finally, within the realm of temporal Graph Convoluational Neural Networks there has been limited work applying temporal GCNs to urban environments. The most similar application is that of traffic prediction in which a connected network of roads is used to predict traffic on roads at future time-steps ~\cite{TGCNTrafficZhao}. However, there is a gap in the literature in the application of temporal GCNs beyond traffic prediction. This is where the primary novelty of this present paper lies. It is the first work using dynamic temporal GCNs to predict urban demand. It shares a novel methodology using both spatial and topological features to inform dynamic predictions.

%% file: 4_dataset.tex
Within the last decade, Online Location-based Services have experienced a surge in popularity, attracting hundreds of millions of users worldwide. These systems have created troves of data which describe, at a fine spatio-temporal granularity, the ways in which users visit different businesses and areas of a city. We hypothesize these data can be used to build a predictive model of the future popularity of a business. To this end, we utilize data from Foursquare, a location technology platform with a consumer application that allows users to check into different locations. As of August 2015, Foursquare had more than 50 million active users and more than 10 billion check-ins ~\cite{foursquarenumbers}.

The basis of our analysis is a longitudinal dataset from two cities that spans three years, from 2011 to 2013, and included over 1.5 million checkins.  
For each venue, we have the following information: geographic coordinates, specific and general category, creation date, total number of check-ins, and number of unique visitors. The specific and general categories fall within Foursquare’s API of hierarchical categories. A full list of the categories can be found by querying the Foursquare API \cite{venueapi}.
The dataset also contains a list of transitions within a given city. A transition is defined as a pair of check-ins by an anonymous user to two different venues within the span of three hours. It is identified by a start time, end time, source venue, and destination venue.
 We consider the set of venues $V$ in a city. A venue $v \in V$ is represented with a tuple $<loc, date, category>$ where $loc$ is the geographic coordinates of the venue, $date$ is its creation date, and $category$ is the specific category of the venue.
The creation date, $date$, for a given venue refers to the date it was added to the Foursquare platform. 

%% file: 5_urbanactivitynets.tex
We begin by examining transitions between Foursquare venues of different category types that we refer to as urban activities. While we are considering a mix of categories users check in in the city, our focus from an analysis and modelling point of view will be focusing on urban activities corresponding to retail establishments (e.g. restaurants). 
In summary, in the following paragraphs we note there exists visible structure in the network of urban activities which varies spatially from city to city as well as temporally across different times of days (e.g. morning versus evenings). We first visualize these trends and then quantitatively measure differences in their structure. We focus London and Paris for our analysis because they are large metropolitan cities with a high Foursquare user base. However, our architecture and model can be applied to any city.
\begin{table*}
\centering
\captionsetup{justification=centering}
\begin{tabular}{>{\centering\arraybackslash}m{1.6cm}|>{\centering\arraybackslash}m{1.1cm}>{\centering\arraybackslash}m{1.1cm}>{\centering\arraybackslash}m{1.1cm}>{\centering\arraybackslash}m{1.1cm}|>{\centering\arraybackslash}m{1.1cm}>{\centering\arraybackslash}m{1.1cm}>{\centering\arraybackslash}m{1.1cm}>{\centering\arraybackslash}m{1.1cm}}
    \toprule 
 \multicolumn{5}{c}{\qquad\qquad\qquad\quad \textbf{London}}     & \multicolumn{4}{c}{\textbf{Paris}}      \\  \toprule 
& AM & Random AM & PM & Random PM & AM & Random AM & PM & Random PM \\ \hline
\# of nodes & 176 & 176 & 199  & 199 & 157 & 157 & 150 & 150  \\ \hline
\# of edges & 1727 & 1125 & 2539 & 1636 & 1708 & 1084 & 1818 & 1105  \\ \hline
$<C>$  & 0.655   &  0.361         & 0.657   & 0.410  & 0.671   &      0.456     &  0.701  &   0.437       \\ \hline
$<C_c>$ & 0.298   &  0.426         &   0.373 &   0.444        &  0.367  &   0.445        & 0.390    & 0.454          \\ \hline
$Q$ &  0.380  &   0.172        &  0.398  &    0.153       & 0.326   &  0.156         & 0.310   &   0.147        \\ 
\bottomrule
\end{tabular} 
\caption{Network metrics for London and Paris for during the morning AM (6am - 12pm) and evening PM (6pm-12am). These metrics are compared to a configuration model (Random).}
\label{tab:networkMetrics}
\end{table*}

\subsection{Visualizing Mobility Interactions}
To visualize an urban activity network, we create a graph $G_i$ for each city $i$, where the set of nodes $N_{cat}$ is the set of business categories defined previously in Section \ref{sec:data}. In this network, business categories are linked by weighted directed edges $e_{s\rightarrow d}$. A directed link is created from the source category $c_s$ to the destination category $c_d$ if at least one transition happens during the time window we consider (e.g. weekend, weekday, or a period of hours during a day). Thus, the weight of each edge is proportional to the total number of transitions from the source category to the destination category for the particular time period of interest for each city. The weights are then normalized by the total number of check-ins that occurred at $c_d$. Therefore, the weight can be interpreted as the percentage of customers of $c_d$ who come from $c_s$. To eliminate insignificant links, we filter out edges that have less than 50 transitions total. We examine two time intervals of interest: morning AM (6am-12pm) and evening PM (6pm-12am). 

In Figure~\ref{fig:networkViz} we visualize the network in the evening for two cities, London and Paris. The colors represent different communities, obtained using the Louvain community detection algorithm~\cite{blondel_2008}. Further, the size of nodes is proportional to their degree. This visualization, as one example, describes similarities and variations in the structure of urban activities in different cities.  We observe an underlying common structure for the two cities, even though cultural distinctions can also be noted. We have observed a similar pattern across different cities which we don't visualize due to lack of space. In terms of similarity in network structure, we see a shopping cluster (green) centered around Department Stores;  a cluster for travel and transport (blue) centered around categories such as Train Stations and Subways; a leisure cluster (light brown) centered around Plazas and containing outdoor categories (e.g. Parks, Gardens, Soccer Stadiums). On the other hand, differences in network structure become also apparent. We note for instance how recreation activities in the evenings differs across the two cities. London has a considerably large nightlife cluster (red) centered around pubs from which a number of different nightlife categories unfold (e.g. Nightclubs, restaurants of different types, Theater). Paris is more segregated and contains two nightlife clusters: one cluster around French Restaurants (red) linked to Coffee Shops, Theaters, Nightclubs; and another cluster (gray) centered around Bars which contains Food Trucks, Fast Food Restaurants, and Music Venues. This dichotomy translates to the presence of two classes of customers each of which adheres to different types of activity sequences during nighttime. Another observation is regarding variations in network structure over time: the Coffee Shop category in London is separated from the nightlife cluster, which may indicate different kind of customer behaviors between daytime and evening. Interestingly, we also see associations emerging between types of businesses. Taking Paris as an example, French Restaurants interact a lot with Coffee Shops and Nightclubs and so do Bars with Food Trucks. In both cities, Coffee Shops are drawing crowds from Subways, Toy Stores with Electronics Stores and Sport Stores.

Overall, these results suggest strong structural characteristics in urban activity networks where different categories of places form interaction patterns of cooperation, where mobile users move from one to the other. Competition on the other hand manifests in a more implicit manner in the network in two ways: first, retail facilities that are grouped in the same node (e.g. Bars) have to share customers that have been previously performing a different activity (e.g. going to a Restaurant) and second, through activities that do not share an edge in the network and as a result they do not interact with one another in terms of mobility patterns. 

\subsection{Network Properties}
\label{sec:network_prop}
We next quantify the structure of these networks in terms of different network properties considering also different time intervals. For our two cities of comparison, we list the network metrics in Table~\ref{tab:networkMetrics} and enlist those next.\\
\begin{itemize}
    \item \emph{The average clustering coefficient}, $<C>$, is the tendency of categories to form triangles, that is to gather locally into fully connected groups. It varies between 0 and 1 with higher values implying a higher number of triangles in the network (see \cite{newman_2010} for more details). 
    \item \emph{The average closeness centrality}, $<C_c>$, is the average length of the shortest path between the nodes and here accounts for the tendency of categories to be close to each in terms of shortest paths \cite{newman_2010}. It varies between 0 and 1 where a higher closeness centrality score for a node suggests higher proximity to other nodes in the network.
    \item \emph{The modularity}, $Q$, is a well established metric indicating how well defined communities are within the network \cite{blondel_2008}. Modularity values fall within the range $[-1, 1]$, with greater positive values indicating greater presence of community structure.
\end{itemize}  

We compared our network metrics to 3 random baselines: an Erd\"os-R\'enyi model\cite{Erdos_1959}, a Barab\'asi-Albert\cite{Albert_statisticalmechanics} model and a configuration model\cite{molloy_1995} which maintains the same degree distribution. Table \ref{tab:networkMetrics} shows the comparison with the configuration model. However, similar results and significance were observed with the two other random models. This comparison provides an indication of how significant empirical observations are with respect to random case. First we note that for all three metrics the real networks are very different to the corresponding null models. 
In general, high clustering coefficient and modularity together with a lower closeness centrality scores point to the tendency of local businesses to form significantly tight clusters that are well isolated from one another. Furthermore, we also investigated variations of these networks properties for different period of the day. We observed in some cases that the closeness centrality was higher in the evening relative to morning hours, which is the case of London for example (see Table \ref{tab:networkMetrics}).

Looking closer at the network modularity scores presented in Table \ref{tab:networkMetrics} we note a partitioning of different categories into communities with scores around $0.3/0.4$ for both cities compared to much smaller values $\approx0.15$ for the null model.
Finally, the similarity in terms of network properties values between the two cities, as well as the prominent community structure in both suggest that the hypothesis that the organization of the retail business ecosystem is similar across cities is a plausible one. This is true to a certain degree, nonetheless variations are also noted due to apparent cultural differences. 

In this section, we highlighted the dominance of community structure and local clustering in urban activity networks. This observation suggests that categories gain (or lose) attractiveness as a result of other activities around them. It further raises the question of how businesses affect each other. In the next section, we look at the evolution of number of checkins and we introduce our deep learning architecture to model the demand of venues over time to shed light on the aforementioned questions. 

\begin{figure}
    \centering
    \begin{subfigure}{1.0\columnwidth}
    \includegraphics[width=1.0\columnwidth]{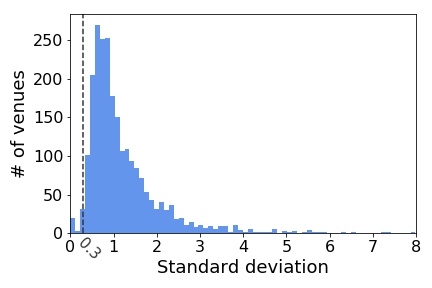}
    \end{subfigure}
    \begin{subfigure}{1.0\columnwidth}
    \includegraphics[width=1.0\columnwidth]{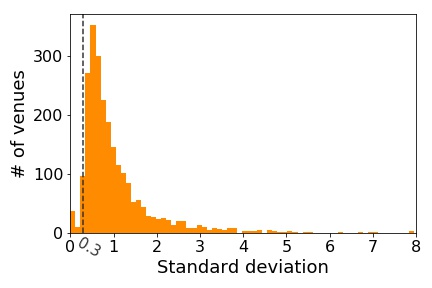}
    \end{subfigure}
        \caption{Histogram of the standard deviation of the relative change in checkins for Paris (blue) and London (orange). For each venue we computed the monthly change in checkins relative to the prior month. For instance, a venue doubling its number of checkins from a month to another would be a change of 1. This gave for each venue a sequence of changes. We then computed the standard deviation of each sequence to evaluate the amplitude of changes.}
    \label{fig:std_changes}
\end{figure}

\subsection{Temporal Trends}
\label{sec:temporal}
In this section, we investigate how the number of venue checkins changes from month to month. To do so, for each venue we compute relative change from one month to the next and calculate the standard deviation (STD) of the sequence of changes. For example, given a venue with demand data over a 12 month time-span. This data contains a sequence of 11 changes in demand. For each change, we calculate the relative change in the number of checkins ($(C_{t+1}-C_{t})/C_{t}$, where $C_t$ is the number of checkins at time $t$). If we assumed that the number of checkins remains constant or had very few changes from month to month this would return a low standard deviation of the relative changes. However, Figure \ref{fig:std_changes} rejects this assumption. Figure \ref{fig:std_changes} shows the histogram of sequence standard deviations for the city of Paris (in blue) and London (in orange). A venue in the city corresponds to a unit point in the histogram. In order to remove noise due to venues with low number of checkins, we filtered out venues which did not reach an average of 20 checkins per month. We observe in Figure \ref{fig:std_changes} that for both cities most of venues have an STD over 0.3, with the histogram peak around 0.5. This highlights that individual venues vary significantly from month to month. This motivates our prediction task in the subsequent section.

%% file: 6_model_analysis.tex
We next describe our model and give an overview of our deep learning approach and architecture. Our model incorporates both spatial and topological features into an enhanced representation which is then fed into a temporal model to predict the next time step. This more complex representation includes attributes of both the spatial and topological domain which enables better predictions. Below, we give an overview of GCNs and LSTMs and describe how our architecture incorporates both.
 \begin{figure*}
\centering
   \centering
   \makebox[\textwidth][c]{\includegraphics[width=18cm]{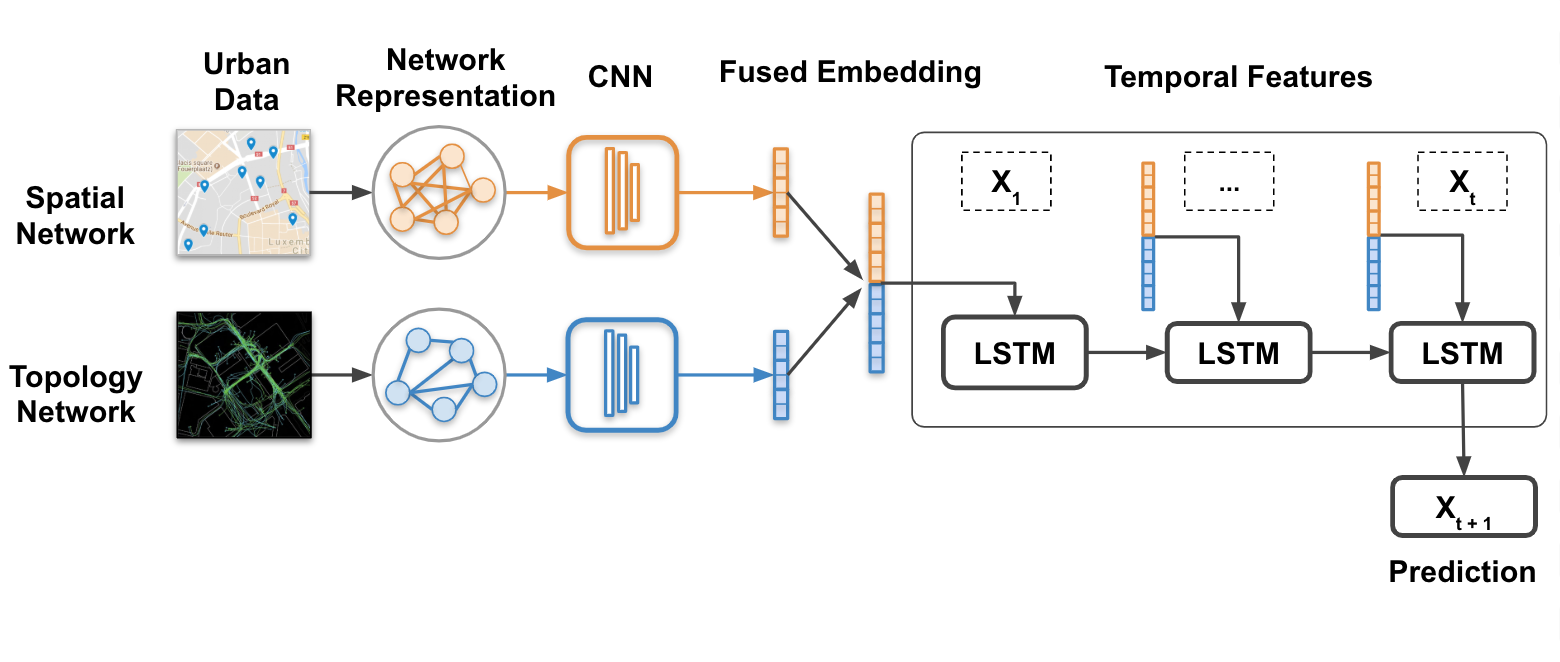}}
    \caption{An overview of the architecture to build spatial and topological representation of the venue graph which is then fed into a temporal model. Fed into the LSTM is the input and the previous hidden state.}
    \label{fig:architecture}
\end{figure*} 

\subsection{Graph Convolution Networks}
Traditional convolutional neural networks (CNNs) are used to represent spatial relationships in Euclidean space for applications to 2D matrices, grid-like structures, or images, amongst others \cite{CNN1_image}. Recent research has worked to extend the principles of a convolutional filter to other applications including graph-structured data. A graph convolutional network (GCN) is a generalization of a CNN and can be applied to capture the topological relationships present between venues urban environments. GCNs harness spectral graph theory by using a spectral filter which is based on the graph Laplacian matrix. Spectral-based GCNs have been increasingly used in numerous applications that involve graph-structured data. These application include image classification, population-based disease predictions, and recommendation systems \cite{spectralImage, spectralDisease, spectralRecommendation}. To date, the only application of GCNs to urban environments has been that of traffic prediction. \cite{TGCNTrafficZhao}. These works represent the network of connected streets in a city as a set of connected nodes in a graph. However, in addition to traffic prediction, GCNs have tremendous potential to model the dynamics and connectivity of venues in urban environments. In this application, the spectral filter captures network features of nodes in the graph as well as in their neighborhood. A GCN model is then built by stacking convolutional layers. For nodes in our network, we define a k-th order neighborhood of a venue $v_i$ as follows: 
\begin{equation}
NB_i = \{v_i \in V \enspace | \enspace d(v_i, v_j) \leq k \}
\end{equation}
where $d(v_i, v_j)$ is the distance in hops from venue $v_i$ to venue $v_j$. A one-hop neighborhood is defined as the adjacency matrix. Further, the k-th hop adjacency matrix is equal to the k-th product of A. The k-th order adjacency matrix is defined as follows:
\begin{equation}
   A^k = \prod_{i=1}^{k} A + I 
\end{equation}
A 1-hop graph convolution is defined as follows:
\begin{equation}
    f(X, A) = \hat{A} X  W
\end{equation}
Where X is our feature matrix, W is the weight matrix and 
\begin{equation}
\hat{A} = \widetilde{D}^{-\frac{1}{2}}A\widetilde{D}^{-\frac{1}{2}}
\end{equation}
Where $\widetilde{D}$ is the degree matrix. We employ two GCNs which build spatial and topological characteristics respectively. Spatial features represent the location of venues relative to others and topological features represent the connectivity of a venue to other venues. These are then fed into a long short-term memory (LSTM) network which is described in detail next. 

\subsection{LSTM}
At each time-step, the network represented is fed into an LSTM which acquires the dynamics dependence and trends over time \cite{HochreiterLSTM}. Generally, an LSTM unit is composed of a cell, an input gate, an output gate, and a forget gate. The cell remembers values of a set time interval and the gates determine the flow of information into the cell and out into the subsequent cell. LSTM models have widely been used on time series data cross numerous domains. Within the context of urban environments, LSTMs have been used to model population mobility flow, bicycle availability, and traffic congestion. Very few works have used LSTMs in conjunction with GCNs. A previous work by Zhao et. al ~\cite{TGCNTrafficZhao} built a deep learning model which fed a GCN into a gated recurrent unit (GRU) to better predict traffic on roads in a city. GRUs are very similar to LSTMs in their architecture. Beyond this work by Zhao et. al, there a notable gap in the literature in the application of temporal GCNs for urban applications. 

Our model aims to predict the number of check-ins to a venue at the subsequent time step. We showed above in Section~\ref{sec:temporal} that the number of check-ins varies from month to month at the venue level and is not consistently stable. We make the assumption that these fluctuations from month to month are based on numerous factors including seasonality, neighborhood growth, and real-time events. These fluctuations could also be an artefact of our dataset which may not represent a significant sample of users. However, as we see below our results suggest that our features capture some variance. These trends suggest that a prediction task for demand must incorporate data from not only previous time-steps but also from other features such as the neighborhood of a venue. We suggest that an LSTM is capable of incorporating these attributes as it is able to remember previous trends of venue demand when trained on data that incorporates numerous features relating to the venue of interest. 
  
\subsection{Model Architecture} 
The aim of our model is to predict the demand of a given venue at the subsequent month. Our analysis below focuses on the cities of London and Paris however the methodology can be applied more broadly to other cities worldwide. For our prediction task, we use check-ins to a venue as a proxy for the popularity of that venue. In Section~\ref{sec:network_prop}, we built a graph for each city of categories to characterize relationships between them. In this section, we build a network of venues, each venue is represented by a node. An edge is drawn between two nodes if at least one transition occurs between the pair of venues. We hypothesize that venue demand is a nonlinear combination of three sets of attributes: physical location in space, venue connectivity, and temporal trends. Our model incorporates and fuses spatial and topological data which are then used to train a dynamic model. Spatial data represents the physical location of venues and their distance between each other. Topological data represents the interactivity and interconnectedness of different venues. We describe both in more detail below. \\ \\
\noindent\textbf{Pre-processing:} To eliminate noise and venues that may not have significant use, we set a threshold of at least a mean of 20 check-ins per month. We do not consider venues that do not meet this threshold.

\subsubsection{Topological Graph}
We build a weighted graph $G = (V, E_t)$ to describe the topological structure of venues. The temporal graph consists of a series of graphs over $T$ time steps $G = \{G_1, G_2, ..., G_T\}$. 
The set of nodes $V = {v_1, v_2, ..., v_N}$ represent the set of individual venues in our graph. $E_t$ represents the number of transitions between two given venues at time step $t$. When feeding our graph into our model, we represent our graph with two data structures:  an adjacency matrix and a feature matrix. The input feature matrix is expressed as $X \in \mathbb{R}^{N\times F}$ where $N$ is the number of nodes and $F$ is the number of features for each node. The number of node attribute features is equal to the length of the historical time series. The adjacency matrix $A \in \mathbb{R}^{N\times N}$ represents the connectivity of the graph $G$. The adjacency matrix contains elements of 0 or 1. The value is 1 if there is ever a transition between two venues and 0 otherwise. For our model, each node in our network is a venue and features represent the number of check-ins to each venue at that time step and the adjacency matrix represents whether two venues are connected. We set the order of the adjacency matrix to be $k =1$ by default but vary the order below to examine the impact of adjacency order. This topological graph, built from Foursquare check-ins, gives us an indication of how connected a venue is to other venues in a city. This can help inform future demand of a venue as the neighborhood attributes of a venue often relate directly to it's own characteristics. However, it does not directly consider the physical location of that venue in space. To this end, we also build a spatial graph of the same set of venues $V$. 

\subsubsection{Spatial Graph} 
While the topological graph contains aspects of location, in which venues with a close proximity to each other are likely to have higher rates of movement between the two, it does not consider explicitly the physical location. As such we build a spatial graph $G_s = (V, E_t)$ in which nodes are similarly venues and edges represent the Euclidean distance between the coordinates of a pair of venues. This graph is static as it does not contain temporal features. We calculate the closeness centrality for each node in the graph. As closeness centrality represents how near a given node is to other nodes in the graph, it represents that node's location in physical space. As with the topological graph, we represent our spatial features with two structures: an adjacency matrix and a feature matrix. The input feature matrix is expressed as $X \in \mathbb{R}^{N\times 1}$ where $N$ is the number of nodes. The feature for each venue is it's closeness centrality. The adjacency matrix $A \in \mathbb{R}^{N\times N}$ represents the connectivity of the graph $G_s$. The adjacency matrix contains elements of 0 or 1. The value is 1 if the distance between the two venues was less than 1km. The distance of 1km was determined empirically. This network represents the spatial connectivity of the venues.

%% file: 7_results.tex
\begin{table*}
\centering
\captionsetup{justification=centering}
\begin{tabular}{M{2.0cm} | M{2.0cm} M{2.0cm} | M{2.0cm} M{2.0cm}}
    \toprule 
 \multicolumn{3}{c}{\qquad\qquad\qquad\quad \textbf{London}}     & \multicolumn{2}{c}{\textbf{Paris}}      \\  \toprule 
Model & RMSE & MAE & RMSE & MAE  \\ \hline
ARIMA               & 5.9123 & 37.5049 & 7.4211  & 49.9920       \\\hline
SVR                 & 4.0780 & 33.0480 & 5.5088  & 44.8892       \\ \hline
LSTM                & 2.4504 & 27.0280 & 3.1793  & 34.4486        \\\hline
Topological TGCN    & 2.0731 & 26.9817 & 2.9535  & 34.4081        \\ \hline
Topological \& Spatial TGCN       & 1.7508 & 25.3600 & 2.7387 & 29.8400        \\    
\bottomrule
\end{tabular} 
\caption{Performance comparison of our model relative to baselines in both cities.}
\label{tab:baselineComparison}
\end{table*}

We build our temporal graph using data beginning on January 1 2012 until December 31 2013. Each monthly graph in $G$ is built using the check-ins at that month. The data was split such that 60\% was used as the training set, 20\% was used as the validation set, and 20\% as the testing set. We set the time steps of the input to be 4 months and run a sliding window through the series. The output of the model, which the model aims to predict, is the next subsequent time-step of the input sequence. 

The performance of our model and that of our baselines is evaluated by two metrics: Mean Absolute Error (MAE) and Root Mean Squared Error (RMSE). We train our model for a maximum of 500 epochs, using Adam optimization \cite{optimizerAdam} with an initial learning rate of $10^{−5}$. We employ early stopping patience of 5, where training is stopped if the loss does not decrease by a threshold of 0.00001 for 5 consecutive epochs. We use a hidden layer the same size as the number of nodes in the venue network graph (N = 568 for London and N = 482 for Paris). We set the size of hops in the graph convolution to $K=1$ after experimentation which showed that increasing the value of $K$ decreases accuracy of the results. This suggests that the direct neighborhood of venues play a larger role in their demand dynamics than venues that are more distant. All of our models were implemented using PyTorch 0.3.1. 
 
We compare our model (en-TGC) with the following baseline models:
\begin{enumerate} 
\item{ARIMA: Auto-Regressive Integrated Moving Average \cite{arimaAsteriou}. The ARIMA baseline uses solely historical demand fit into a parametric model to predict future demand.}  
\item{SVR: Support Vector Regression \cite{SmolaSVM}. The SVR trains on historical demand data to construct a hyperplane to learn a relationship between the inputs and outputs. We use a linear kernel.}
\item{LSTM: Long Short-Term Memory recurrent neural network \cite{HochreiterLSTM}. The LSTM network is a deep learning model which trains on past demand data for individual venues.}
\item{Topological GCN: The GCN builds a graphical representation of the network using only Foursquare transitions and uses a spectral filter to learn features.}
\end{enumerate}
 
Table ~\ref{tab:baselineComparison} includes the results of our model relative to other baselines enumerated above in both Paris and London. Our proposed architecture outperforms other models in both RMSE and MAE in both cities. We see that ARIMA has a much higher error rate than machine learning methods, highlighting that venue demand prediction is a complex task which involves multiple factors and that the historical demand trends of a venue are likely not sufficient for future predictions. Our model reduces the RSME by 70.39\% relative to the ARIMA prediction in London and by 63.1\% in Paris. Further, we see that our model performs better than a simple LSTM. Our model reduces the RSME by 28\% relative to an LSTM in London and 13\% in Paris. These trends demonstrate the value of network features learned from the spectral graph convolution. It highlights the power of graphical neural networks to model dynamics of urban environments. We also see a decrease in RSME when integrating spatial features into the model. In London RSME decreases by $\approx$ 15\% and in Paris it decreases by $\approx$ 7\%. This demonstrates the benefit of integrating spatial features into the representation in the deep learning model. Our experiments demonstrate that our model can learn spatio-temporal trends of venue demand and consistently outperform baseline models. To the best of our knowledge, this is the first application of temporal graph convolutional neural networks applied to mode business dynamics. We see similar trends in the decrease in MAE by our models. An MAE of $\approx$ 25 in London and $\approx$ 29 in Paris are low given the difficultly of the prediction task at hand. With a minimum of at least 20 checkins per month, venue monthly checkins in our dataset range from 20 to over 1000. With this context in mind, a MAE of $\approx$ 25 in London and $\approx$ 29 in Paris suggests our model is performing extremely well given the inherent complexity of the task.

\begin{figure}
\centering
   \centering
   \includegraphics[width=8cm]{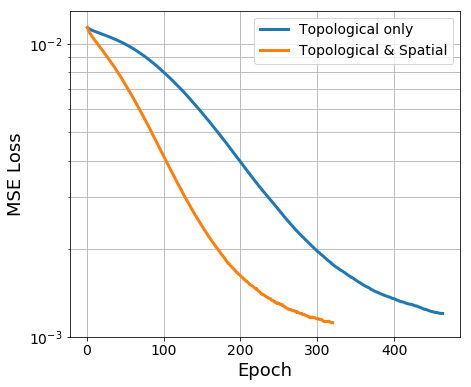} 
    \caption{Validation loss over training epoch for London.}
    \label{fig:mse}
\end{figure}

We next compare the training efficiency of our model with both topological and spatial features to one in which only topological features are used. Figure~\ref{fig:mse} shows the validation loss across training epoch for both models. The max training epoch was set to 500. However, as we employed early stopping in the training process, the number of training epochs are less than 500. Firstly, we see the combined model converges more quickly than the model with only topological features. This suggests that the fusion of both features enables the model to learn more trends more quickly. Additionally, the loss of the combined model decreases more quickly. We see similar trends for loss when training the models in Paris.


In summary, our model with combines spatial and topological features beats all of our baselines reducing the RMSE by up to 28\% relative to a deep-learning baseline. Additionally, our models with both feature types converges more quickly than one without, suggesting it is able to learn trends of venue demand at a faster rate.
 
A limitation of our work is that we only examine the dynamics of businesses that meet our minimum threshold of busyness. While this helps us to focus on only venues with meaningful data, it eliminates venues that may still have interesting trends, online and offline. However, this present study sets the frame for further general studies. Additionally, in this work we conduct a preliminary examination of the variations in network trends across two cities. Future work could expand upon this to explore the duality between general network trends and cultural consumer idiosyncrasies across cities. Our analysis could also be further expanded by considering temporal network analysis, examining the variations in features across different time intervals of interest.  

%% file: 8_conclusion.tex
Our methodology highlights the power of graph convolutional networks in building prediction models within the context of urban cities. This can broadly be applied to many systems in which interactions between agents must be taken into account.
We began in Section~\ref{sec:network} of this work by detecting patterns of cooperation in urban networks and quantifying similarities and differences in network structure across cities. Next in Section~\ref{sec:temporal}, we described the high variability in monthly checkin data, motivating our subsequent prediction task and also illuminating the inherent complexity and challenges in the task.

Using spatial features in conjunction with topological network measures, we developed a temporal machine learning model to predict future business demand in Section~\ref{sec:methodology}. The novelty of our approach in  methodological terms stems from the use of temporal GCNs with a spatial and topological representation of venues in a city to tackle open modern urban questions. As next steps on this work, we plan to incorporate dynamic node network and integrate additional features such as the category of the venue. Further, we plan to incorporate more attributes of the road network of the city, such as traffic congestion, busyness, and speed.

Our model and results can support policy makers, business owners, and urban planners as they have the potential to pave the way for the development of sophisticated models describing urban neighborhoods and predicting future growth trends for venues in a given area. Our methodologies could also urban planners in better understanding conditions for growth for venues and working to determine the optimal conditions for establishing a venue and more generally new urban facilities. Our work is a first step in harnessing dynamic graph convolutional networks to working to model urban dynamics.